\documentclass[12pt]{article}

\usepackage{amsmath}
\usepackage{amssymb}
\usepackage{graphicx}
\usepackage{xcolor}
\usepackage{mathtools}
\newcommand{\review}[1]{\textcolor{black}{#1}}

\DeclarePairedDelimiter{\ceil}{\lceil}{\rceil}

\usepackage{times}

\newenvironment{sciabstract}{%
\begin{quote} \bf}
{\end{quote}}

\title{\vspace{-42.5mm}Crowded transport within networked representations of complex geometries}

\author
{Daniel B. Wilson,$^{1,\dag,\ast}$ Francis. G. Woodhouse,$^{1,\ddag}$ Matthew J. Simpson,$^{2}$\\ Ruth E. Baker $^{1}$\\
\\
\normalsize{$^{1}$Wolfson Centre for Mathematical Biology, Mathematical Institute, University of Oxford, }\\
\normalsize{Radcliffe Observatory Quarter, Oxford OX2 6GG, United Kingdom.}\\
\normalsize{$^{2}$School of Mathematical Sciences, Queensland University of Technology, }\\
\normalsize{Brisbane, Queensland, Australia.}\\
\\
\normalsize{$^\ast$To whom correspondence should be addressed; E-mail: dbwilson@bu.edu.}\\
\normalsize{$^\dag$Present address: Department of Mathematics and Statistics, Boston University, MA 02215, USA.}\\
\normalsize{$^\ddag$Present address: Smith Institute for Industrial Mathematics and Systems Engineering,}\\
\normalsize{ Oxford, United Kingdom.}
}

\date{}

\begin{document} 


\baselineskip24pt


\maketitle


\begin{sciabstract}
Transport in crowded, complex environments occurs across many spatial scales. Geometric restrictions can hinder the motion of individuals and, combined with crowding between individuals, can have drastic effects on global transport phenomena. However, in general, the interplay between crowding and geometry \review{in complex real-life environments} is poorly understood. Existing \review{analytical methodologies are not always readily extendable to heterogeneous environments: in these situations predictions of} crowded transport behaviour within heterogeneous environments rely on computationally intensive mesh-based approaches. Here, we \review{take a different approach by employing} networked representations of complex environments \review{to} provide an efficient framework within which the interactions between networked geometry and crowding can be explored. We demonstrate how the framework can \review{be used to}: extract detailed information at the level of the whole population or an individual within it; identify the topological features of environments that enable accurate prediction of transport phenomena; and, provide \review{insights} into the design of optimal environments. 
\end{sciabstract}

\newpage

\section*{Introduction}


The efficacy of a wide range of cellular processes within living organisms, from protein synthesis \cite{Palade1975_Science} to the initiation of a T-cell immune response \cite{Dushek2012_ImmRev}, hinges upon the timely transport of macromolecules through crowded intracellular environments \cite{Ando2010_PNAS}. The motion of individuals within complex environments is central, but not limited to, cell biology, and is important in a wide range of scientific and technological disciplines. Indeed, understanding the roles that both geometry and crowding play in regulating transport processes has immediate and disparate applications across a vast range of spatial and temporal scales, from designing planning algorithms for autonomous robotic motion \cite{LaValle_CUP06} to utilising the transport of nanoparticles to deliver targeted drug therapies \cite{Nano}. However, despite the ubiquity of applications, a \review{general means to quantify and characterise} the relationships between environmental geometry, crowding and transport phenomena \review{in complex real-life environments} remains elusive. \review{Mathematical studies such as this one can play a key role in identifying and quantifying such relationships.}


The behaviour of crowded individuals has captured the attention of mathematicians and physicists for decades \cite{LizanaPRL08,WeiScience2000,BechingerRevModPhys2016,Benichou2018Review}, and the environments in which the individuals are constrained have been found to have large consequences for the emergent transport behaviour. For particles diffusing along an infinite one-dimensional lattice \cite{Harris1965}, excluded volume interactions between particles significantly hinders the motion of tracer particles, and can induce a qualitative shift in the time-dependence of the mean squared displacement (MSD) from classically diffusive (MSD $\sim t$) to anomalously diffusive (MSD $\sim t^{\alpha}$, where $\alpha<1$). However, for some fractal environments, including diffusion-limited aggregates, the MSD for a tracer particle has the same exponent both in the presence and absence of crowding \cite{Amitrano1985}. Moreover, on comb lattices (one-dimensional backbones with periodic and infinite extrusions representing low-dimensional environments) exclusion interactions along the backbone result in a speed-up of the transport of a tracer particle, and even results in super-diffusive behaviour (MSD $\sim t^{\alpha}$, where $\alpha>1$) on intermediate timescales \cite{BenichouPRL2015}. Crowding in complex environments has also been studied on higher-dimensional (greater than or equal to two) regular lattices, such as diffusion in the presence of obstacles \cite{Ellery2016,LeitmannPRL2013,LeitmannPRE2018} and crowded transport on Manhattan lattices \cite{NJP2020} where environmental complexity arises via disordered directionality in the lattice. The above studies make analytical progress in understanding crowded transport behaviour by utilising symmetry and scaling arguments, as well as exploiting the infinite size of these domains. \review{Whilst less common, there are studies that make analytical progress in understanding crowded transport in finite heterogeneous environments. For example the transport of proteins within heterochromatin has been studied using results from transport within fractal environments \cite{EMBO09}. Additionally, analytical results from lattice gas cellular automata models have been used to study cell migration in heterogeneous environments, by coupling the automata to a force field representing environmental complexity \cite{Hatzikirou08}. However, when interested in exploring microscopic properties of crowded transport in detailed non-fractal real-life environments}, such as the intracellular environment, \review{many of these} analytical approaches are not \review{readily extendable. In such situations,} sophisticated numerical approaches are employed \review{instead.}


Significant technological advances in imaging techniques, such as light-sheet microscopy \cite{TomerLightSheet} and x-ray tomography \cite{DoTomography}, have drastically enhanced the availability of high-resolution images of complex environments (see Fig.~1A). Incorporating these \review{detailed} geometries into mathematical studies of crowded transport requires discretisation of the geometry into a high-resolution mesh, this is a collection of interconnected voxels upon which transport models can be studied numerically. However, numerically integrating equations of motion on these meshes can incur significant computational costs, particularly for stochastic modelling paradigms that require repeated simulations to explore expected transport behaviour \cite{Engblom2009_SIAM_JSC,Englbom2018_RSOS,Isaacson2013_JCP}. If these computational costs are too high \review{they present} a barrier to studying a large ensemble of geometries, as is necessary to \review{properly} characterise how geometry and crowding influences transport behaviour. In addition, it remains unclear how to relate such high-dimensional descriptions of the geometry with statistics of the transport process \review{in order} to fully understand the relationship between geometry and crowded transport.

\begin{figure*}[t]
\centering
\includegraphics[width=\columnwidth]{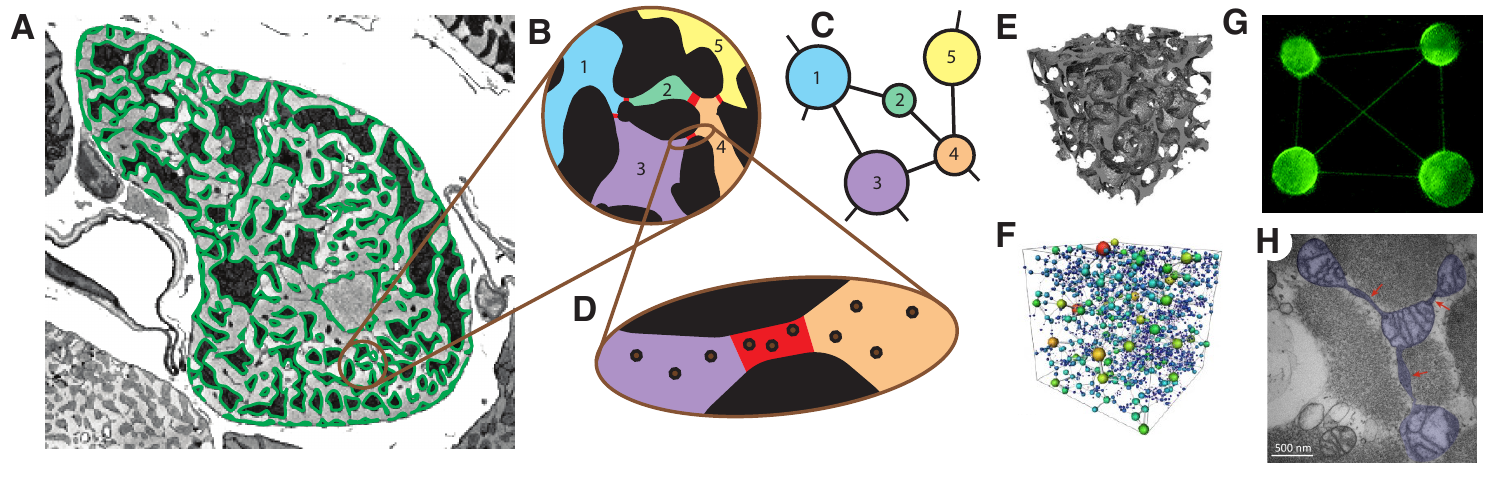}
\caption{Complex environments and their networked representations. (A) An electron-microscopy image highlighting the substructures within a cardiomyocyte cell. The highlighted dark gray regions are mitochondria which act as barriers to macromolecular transport. This image was provided by the Cell Structure and Mechanobiology Laboratory at the University of Melbourne. (B) The free space within a subsection of the cardiomyocyte is segmented into five distinct reservoirs, connected by narrow channels highlighted in red. (C) The natural networked topology arising from the cell segmentation. (D) Macromolecular crowding effects are most prominent within the narrow channels. (E-F) Micro-CT scan of the pore space in a Berea sandstone sample, and the corresponding pore network reproduced from \cite{Blunt2013_AWR}. (G) Nanotube vesicle network reproduced from \cite{Karlsson2002_PNAS}. (H) An electron-microscopy image of three nanotunnels connecting four mitochondria reproduced from \cite{Vincent2017_TrendsCell}. }
\label{Fig1}
\end{figure*}


To circumvent the limitations of high-resolution meshes, we require a framework for studying crowded transport in complex environments that avoids traditional spatial discretisation \review{methods}. A proposed alternative description of complex geometries is to represent the geometry as a network \cite{Blunt2002_AWR,DongPRE09,SNOW}. In particular geologists have used these networked representations to study the transport of fluid and sediment through porous media \cite{Blunt2013_AWR}. There are several available algorithms (such as the maximal ball algorithm \cite{DongPRE09} and SNOW \cite{SNOW}) to extract networked representations that initially dissect complex geometries into several reservoirs (see Fig.~1B). Connecting these reservoirs are narrow regions of space that are referred to as throats. These throats represent highly restricted regions where individuals experience strong crowding, as seen within heterogeneous porous media \cite{Blunt2013_AWR} (Fig. 1E,F), nanotubes in microfluidic devices \cite{Karlsson2002_PNAS} (Fig. 1G), and in newly discovered nanotunnels that connect mitochondria \cite{Vincent2017_TrendsCell} (Fig.~1H). Networks are formed by assigning the reservoirs to be the network nodes, and an edge connects two nodes if the corresponding reservoirs are connected by a throat. The reservoirs are often described as balls with volume equal to the volume of the reservoir, this leads to what is referred to as a balls and sticks network (Fig.~1C).  By virtue of the fact that the reservoirs are typically much larger than a voxel in a mesh reconstruction (Fig.~1C), such networks provide a low-dimensional, efficient characterisation of the complex geometry. Indeed, application of a network extraction algorithm (SNOW \cite{SNOW}) to the cardiomyocyte image in Fig. 1A produces a network with 81 reservoirs, whereas the natural Cartesian mesh (where each pixel is represented as a voxel) has 16774 voxels (SI Numerical Methods Section 4.6, Fig.~S10).


Beyond dimensionality reduction, networked representations allow us to characterise the role of geometry in regulating transport processes due to the breadth of available topological descriptors that can predict networked transport behaviour \cite{Wilson2018_PRE_A,Wilson2019_SIAM_JAM,Woodhouse2016_PNAS}. For example, two classes of summary statistics that are known to accurately predict various emergent transport behaviours are degree-based statistics (these depend solely on the degree distribution of the network, where the degree of a node is equal to the number of edges adjacent to that node) and spectral-based statistics (quantities based on the eigendecomposition of network related matrices, such as the graph Laplacian \cite{Masuda2017}). 

Whilst the majority of the literature on networked transport focusses on single non-crowded individuals, there have been several studies exploring crowded transport processes on networks. The Totally Asymmetric Simple Exclusion Process (TASEP) is a transport process where particles can move only in one direction along one-dimensional segments and individuals cannot bypass each other. The TASEP has been studied extensively on the one-dimensional line \cite{Spitzer1970}, small graphs \cite{EmbleyPRE2009,RaguinPRE2013} and large networks \cite{IzaakPRL2011,ShenChaos2020,BaekPRE2014}, where the uni-directionality of the transport results in various behaviours such as shocks and traffic jams \cite{PinkoviezkyPRE2014}. In addition the TASEP has been used to model\review{, for example,} protein dynamics along microtubule networks \cite{Neri_NJP, Neri_PRL}. In these models proteins are assumed globally well-mixed within a single large reservoir which is coupled to a filament network upon which the proteins perform asymmetric random walks with exclusion.

In contrast, in this work we are interested in the unbiased transport of individuals that are locally well-mixed within the reservoirs that make up the  network. Moreover, the individuals are purely constrained to the network and experience crowding only within geometrically constrained regions (Fig.~1D). As such we build a mathematical framework to study the transport of individuals on general finite networks (in the form of balls and sticks, Fig.~1D) where crowding effects are present only along the edges of the network. We note that there are significantly fewer results concerning symmetric crowded transport on networks compared with the expansive TASEP literature.


We introduce our framework for crowded, networked transport by presenting a hierarchy of diffusive transport models with increasing computational scalability (SI Table 1 provides a summary). The ability of our framework to identify governing principles connecting transport, crowding and geometry is demonstrated through an examination of how crowding and topology affect networked equilibration times \cite{Delvenne2015_NatComms}. Understanding this relationship is crucial as many results within statistical physics assume an equilibrated population, but the validity of this assumption is rarely addressed \cite{Penington16}. A key result of our work is that heterogeneity in the microscopic structure of complex environments enables low-connectivity networks, as seen in networked descriptions of real-world environments (Fig. 1A–D), to achieve globally minimal equilibration times. We conclude by extending our framework to provide information on the dynamics of a single motile individual, an extension which opens the door to studying the dynamics of intracellular signalling pathways \cite{Isaacson2011_PNAS} where the dynamics of individual proteins are known to control a vast range of biological processes, such as T-cell activation and stem cell differentiation.

\section*{Results}
\paragraph*{Individual crowding combined with geometry-induced crowding drastically slows equilibration}

A complex environment (Fig.~1A--D) is described by a network $\mathcal{G} = \{ \mathcal{V} , \mathcal{E} \}$, where $\mathcal{V}$ is the set of reservoirs and their connectivity is specified by the set of edges $\mathcal{E}$. Each edge represents a narrow channel within the geometry where crowding between individuals is non-negligible (Fig.~1D). To incorporate crowding, a narrow channel $(i,j) \in \mathcal{E}$ connecting distinct reservoirs $i$ and $j$ is discretised using a one-dimensional lattice with integer length $K_{(i,j)}$, where individuals undergo a symmetric random walk and at most one individual may occupy each lattice site (SI Extensions Section~$3.1$ provides a relaxation of this assumption), this process is known as a Symmetric Simple Exclusion Process (SSEP) ~\cite{LiggetBook}. Crowded transport within this networked environment is modelled using a canonical framework for stochastic processes, the continuous-time Markov chain (CTMC). A population of $N$ individuals is distributed on the network and their positions evolve as follows. An individual within a narrow channel lattice site attempts to jump into an adjacent lattice site or reservoir at rate $\alpha$. If the adjacent site is already occupied a collision event occurs and the jump is aborted (Fig.~2D). However, due to the volumetric differences between reservoirs and narrow channels (Fig.~1B) crowding effects in reservoirs are assumed negligible and so jumps into reservoirs are never aborted. Individuals within each reservoir are assumed well-mixed. Those in reservoir $i$ attempt to jump into the first lattice site of one of their connecting narrow channels at rate $\gamma_i$ (Fig.~2D), this means that $\tau_i = \gamma_i^{-1}$ is the average time taken for an individual in the $i$-th reservoir to exit the reservoir. This exit time depends strongly on the local geometry of each reservoir \cite{Markowsky2011_ECP,Benichou2008_PRL} and can be calculated as follows. To calculate the mean reservoir exit time for a well-mixed individual we can first consider narrow exit time problems, for which both analytical and computational approaches are readily available \cite{Benichou2008_PRL,Condamin2005_PRL,Herrmann2016_SIAM_JSC}. Narrow exit times  quantify the time taken for a particle to reach a narrow opening (such as the opening of a narrow channel) conditional on the initial position of the individual. Integrating these quantities over the reservoir and normalising by the reservoir volume, will yield mean exit times for well-mixed individuals. However, in the interest of maintaining generality we keep the parameters $\tau_i$ to be abstract, but note particular reservoir geometries can be incorporated as described above. This CTMC is referred to as the full Markov model (FMM) and a technical description of the FMM is found in SI Models Section~$1.1$.


Evaluating how geometry combined with crowding can impact transport behaviour requires a suitable summary statistic with which to characterise transport. Intuitively, it can be reasoned that the effects of crowding between individuals along narrow channels should result in a ``slowing down" of the transport process as individuals in the channels block the paths of others. The insightful statistic to quantify such an effect is the time taken for an initially unequilibrated population of individuals to become well-mixed. This time is known as the equilibration time and is calculated as the reciprocal of the spectral gap (second smallest eigenvalue in absolute value) of the transition matrix for the CTMC \cite{Chung1996_Book}. The average reservoir mean exit time, $\langle \vec{\tau} \rangle = \left( \sum_{i=1}^{|\mathcal{V}|} \tau_i \right) / |\mathcal{V}|$, quantifies the time for an individual to attempt to leave the average reservoir. As the individual $\tau_i$ are dictated by the geometry of each reservoir, the average reservoir mean exit time is a global geometric descriptor of the complex environment\footnote{Considering higher order moments of the reservoir exit time would reveal higher order geometric effects, however these cannot be incorporated in a Markovian model such as the one we consider here.}. By computing the spectral gap of the transition matrix of the FMM, we explore the relationship between the equilibration time and the average reservoir exit time for a small network (Fig.~2A, inset) in both the absence and presence of crowding. In the absence of crowding, networks with lower average reservoir mean exit times will facilitate quicker equilibration (Fig.~2A, dot-dashed curve). Intuitively this is unsurprising as individuals attempt to move between reservoirs more frequently.  With crowding effects incorporated, the monotonicity between the average reservoir mean exit time and the equilibration time is lost (Fig.~2A, solid curve). The equilibration time rises once the volume-excluding interactions within the narrow channels induce a bottle-neck that impedes transport between reservoirs. Thus, a combination of the localised geometry responsible for creating narrow channels (geometry-induced crowding) and interactions between the individuals within a narrow channel (crowding between individuals) can significantly increase the time taken for a population to equilibrate. \review{We derive an analytical condition for when the equilibrium occupancy of each narrow channel lattice site is high and thus crowding effects are important}, which is $\langle \vec{\tau} \rangle \ll  N_{\text{HD}} / \left( \alpha |\mathcal{V}| \right)$, where $N_{\text{HD}} = \left( N - K_{\text{tot}} \right)$  is the effective number of particles in the reservoirs and $K_{\text{tot}}$ is the total number of lattice sites across all narrow channels (SI Models Section~$1.1.3$). We will term this regime the high-density regime and it determines when the combination of localised geometry and crowding is most prominent (Fig.~2A, left shaded region). 

Whilst the FMM is a conceptually ideal way to describe crowded transport in complex geometries, evaluating the equilibration time is computationally infeasible for all but the simplest networks (Fig.~2A, inset) as the dimensionality of the transition matrix for the FMM is $\mathcal{O} \left( 2^{K_{\text{tot}}} N^{|\mathcal{V}|-1} \right)$ (SI Models Section~$1.1.1$). Therefore, to further investigate the role of crowding and geometry on transport behaviour it is critical to consider dimensionality reduction techniques.

\begin{figure}
\centering
\includegraphics[width=11.7cm]{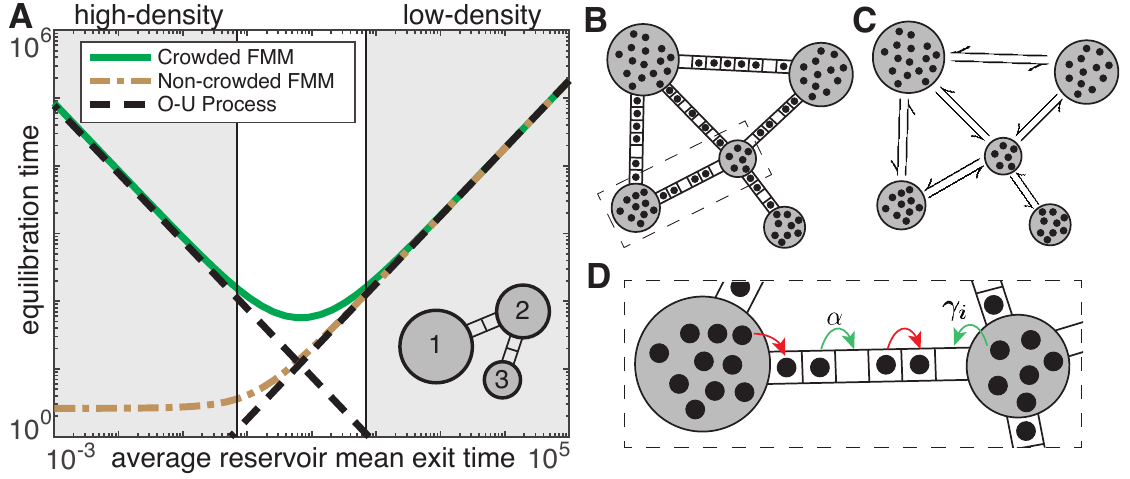}
\caption{ A combination of crowding and geometry can impede population level transport. (A) Equilibration time of the FMM with individual crowding (solid curve) and without (dot-dashed curve) for a three-reservoir network (inset). The equilibration time of the Ornstein-Uhlenbeck process in both the low- and high-density regimes (dashed black curves). The low- and high-density regimes are highlighted in grey for $ \langle \vec{\tau} \rangle \geq 10 \left( N - K_{\text{tot}} \right) / \alpha |\mathcal{V}|$ and $ \langle \vec{\tau} \rangle \leq 0.1 \left( N - K_{\text{tot}} \right) / \alpha |\mathcal{V}|$, respectively. The parameters are $K_{(1,2)} = K_{(2,3)}=2$, $\vec{\tau} = \delta \left( 2,4/3,1\right)$ such that $\langle \vec{\tau} \rangle = 13 \delta / 9$ and $\langle \vec{\tau} \rangle$ varies from $10^{-3}$ to $10^{5}$, $\alpha=1$, and $N=25$. (B) Diagrammatic representation of the FMM where narrow channels are discretised as integer lattices along which individuals undergo crowded transport. (C) Diagrammatic representation of the RMM where individuals hop directly between adjacent reservoirs. (D) Diagrammatic representation of the crowding interactions seen within the FMM.}
\label{Fig2}
\end{figure}

\paragraph*{Scaling to large geometries}

Networked representations of complex environments (Fig.~1A) may contain on the order of hundreds or even thousands of interconnected reservoirs. In light of such, we require models that scale computationally to large environments whilst incorporating details of their microscopic spatial structure. The high dimensionality of the FMM arises from explicitly modelling the occupancy of every lattice site along every narrow channel. 

To make progress we introduce a reduced Markov model (RMM) which, in lieu of considering the dynamics within the narrow channels in detail, allows for direct exchange of individuals between reservoirs (Fig.~2C). This approach is reminiscent of the average current calculations for exclusion processes between two open boundaries \cite{Derrida2004}. For the RMM let $\vec{n}$ be the configuration vector, where $n_i$ is the number of individuals in the $i$-th reservoir. Focussing on the high-density regime (SI Models Section~$1.2.3$ discusses the low-density regime), where crowding effects are most important, the rate at which exchange between two connected reservoirs $i$ and $j$ occurs, denoted $k_{i,j}^{\text{HD}}(\vec{n})$, is calculated by considering the dynamics of the interacting individuals along the narrow channels (SI Models Section~$1.2$). By invoking particle-hole duality \cite{Beijeren1983_PRB}, it is convenient to consider the dynamics of the vacant sites rather than the individuals explicitly. The resulting expression for $k_{i,j}^{\text{HD}}(\vec{n})$ is given by 
\begin{equation}\label{eq:HD_Transition_Rate}
k^{\text{HD}}_{i,j}(\vec{n}) = \dfrac{1}{K_{(i,j)}-1}\left(\dfrac{\tau_j^{-1} (n_j+1) + 2\alpha}{\alpha^2} + \dfrac{K_{(i,j)}-2}{2\alpha} \right)^{-1}.
\end{equation}
Similar to the FMM, the RMM is a CTMC and the reciprocal of the spectral gap of the transition matrix provides the equilibration time. However, the dimensionality of the RMM is still prohibitively high, $\mathcal{O} \left( N^{|\mathcal{V}|-1} \right)$, and does not scale computationally to large complex environments.

Further dimensionality reduction can be achieved through a continuous mean-field approximation of reservoir occupancy in the RMM\footnote{A continuous approximation of reservoir occupancy is only valid if the number of individuals in the RMM is large. Fortunately this is exactly when an approximation of the RMM is necessary.}. Introducing $\vec{x}$, such that $x_i = n_i/N_{\text{HD}}$ is the fraction of the population that lies within the $i$-th reservoir, and expanding the chemical master equations governing the RMM as a Taylor series provides the corresponding Fokker-Planck equation, a partial differential equation describing the evolution of the probability density for the distribution of individuals (SI Models Section~$1.3.1$, Eq.~[S24]). In equilibrium, the networked distribution of the population is determined by the reservoir geometry and is given by $\vec{x}^*$ such that $x_i^* = \tau_i / \sum_{j=1}^{|\mathcal{V}|} \tau_j$ (SI Models Section~$1.3.2$). Localising the Fokker-Planck equation about $\vec{x}^*$ reveals that the population dynamics as the networked distribution equilibrates is governed by an Ornstein-Uhlenbeck (OU) process which is known to be Gaussian \cite{Risken1989_Book}. Thus, from an initial configuration of individuals $\vec{x}_0$, the configuration at time $t$ follows a multivariate normal distribution with known mean vector $\vec{\mu}(t;\vec{x}_0)$ given by
\begin{equation}\label{eq:OU_mean}
\vec{\mu}(t;\vec{x}_0) = \exp \left( -t \mathbf{F}^{\text{HD}} \right) \vec{x}_0 + \left( \mathbf{I} - \exp \left( -t \mathbf{F}^{\text{HD}} \right) \right) \vec{x}^*,
\end{equation}
where $\mathbf{I}$ is the identity matrix, and covariance matrix $\mathbf{\Sigma}(t;\vec{x}_0)$ given by
\begin{equation}\label{eq:OU_covar}
\mathbf{\Sigma}(t;\vec{x}_0) = \int_0^t \exp \left( -u \mathbf{F}^{\text{HD}} \right) \mathbf{D}^{\text{HD}} \exp \left( -u \left( \mathbf{F}^{\text{HD}} \right)^{T} \right) \mathrm{d}u,
\end{equation}
where $\mathbf{F}^{\text{HD}}$ and $\mathbf{D}^{\text{HD}}$ are drift and diffusion matrices, respectively (see SI Models Section~$1.3.3$ for a derivation). The entries of the drift matrix $\mathbf{F}^{\text{HD}}$ are given by
\begin{equation}\label{eq:driftHD}
F^{\text{HD}}_{i,j} =  \mathcal{A}_{i,j} \dfrac{ \left( k_{i,j}^{\text{HD}}\right)^{\prime}(N_{\text{HD}}x_j^*)}{N_{\text{HD}}},
\end{equation}
for $i \neq j$, and $F^{\text{HD}}_{i,i} = - \sum_{j\neq i} F^{\text{HD}}_{j,i} $ for $1 \leq i \leq |\mathcal{V}|$, where $\mathcal{A}_{i,j}$ are the entries of the network adjacency matrix and $k_{i,j}^{\text{HD}} (N_{\text{HD}} x_j^* )$ is the continuous extension of $k_{i,j}^{\text{HD}}(\vec{n})$ defined in Eq.~\eqref{eq:HD_Transition_Rate}. The entries of the diffusion matrix $\mathbf{D}^{\text{HD}}$ are
\begin{equation}
D^{\text{HD}}_{i,j} = -  \mathcal{A}_{i,j}\dfrac{ \left( k^{\text{HD}}_{i,j}(N_{\text{HD}}x_j^*) + k^{\text{HD}}_{j,i}(N_{\text{HD}}x_i^*) \right)}{2 N_{\text{HD}}^2},
\end{equation}
for $i \neq j$, and $D^{\text{HD}}_{i,i} = - \sum_{j \neq i} D^{\text{HD}}_{i,j}$ for $1 \leq i \leq |\mathcal{V}|$. Equation \eqref{eq:OU_mean} reveals that the equilibration time of the OU process is dictated by the spectral gap of the drift matrix $\mathbf{F}^{\text{HD}}$ which is in the form of a networked graph Laplacian. The spectral gap of $\mathbf{F}^{\text{HD}}$ accurately predicts the equilibration time for the FMM (Fig.~2A) and is inexpensive to compute. In particular, the transition matrix for the FMM in Fig.~2A has a dimension of $6968$ whilst the weighted graph Laplacian $\mathbf{F}^{\text{HD}}$ has a dimension of three. We now exploit this remarkable dimensionality reduction to reveal the fundamental principles that govern topological optimisation of equilibration times in crowded environments.

\paragraph*{Equilibration times are highly sensitive to an environments network topology}
\review{To demonstrate how network topology can affect equilibration times we first consider a toy network consisting of five reservoirs (Fig.~3A inset). The three-dimensional reservoir positions $\vec{x}_i$ are uniformly sampled within a unit sphere. The integer narrow channel lengths between reservoirs $i$ and $j$ are given by $K_{i,j} = \ceil{||\vec{x}_i - \vec{x}_j||/\Delta}$, where $\Delta = 2 \times 10^{-3}$ is the width of the narrow channels and $||\vec{x}_i - \vec{x}_j||$ is the Euclidean distance between reservoirs $i$ and $j$. We now consider all $728$ possible connected networks with five reservoirs and their corresponding equilibration times.  }

As a function of topology, equilibration times, calculated from the spectral gap of $\mathbf{F}^{\text{HD}}$, vary over several orders of magnitude (Fig.~3A). The topology that induces the quickest equilibration is the complete network (Fig.~3A,B--(xiv)), because the opportunities for individuals to exchange between connected reservoirs are maximised when every connection is present. However, a complete network is inappropriate to describe most complex environments due to spatial constraints limiting the connectivity of the reservoirs (Fig.~1A--D). The connectivity of a network can be quantified by its total edge length, the sum of the narrow channel lengths present in the network. Imposing a restriction on the total edge length (Fig.~3A, vertical line) reveals a new non-complete optimal network (Fig.~3A,B--(viii)). By varying the restriction over the range of total edge lengths, as defined by the minimum spanning tree(s) and the complete network (Fig.~3A,B--(i) and (xiv), respectively), a frontier of optimal networks arises (Fig.~3A,B). Networks that lie on the optimal frontier, which we term the \textit{optimal networks}, represent environments in which a population equilibrates efficiently, under a given restriction on the environmental connectivity. 

\begin{figure*}[!th]
\centering
\includegraphics[width=\columnwidth]{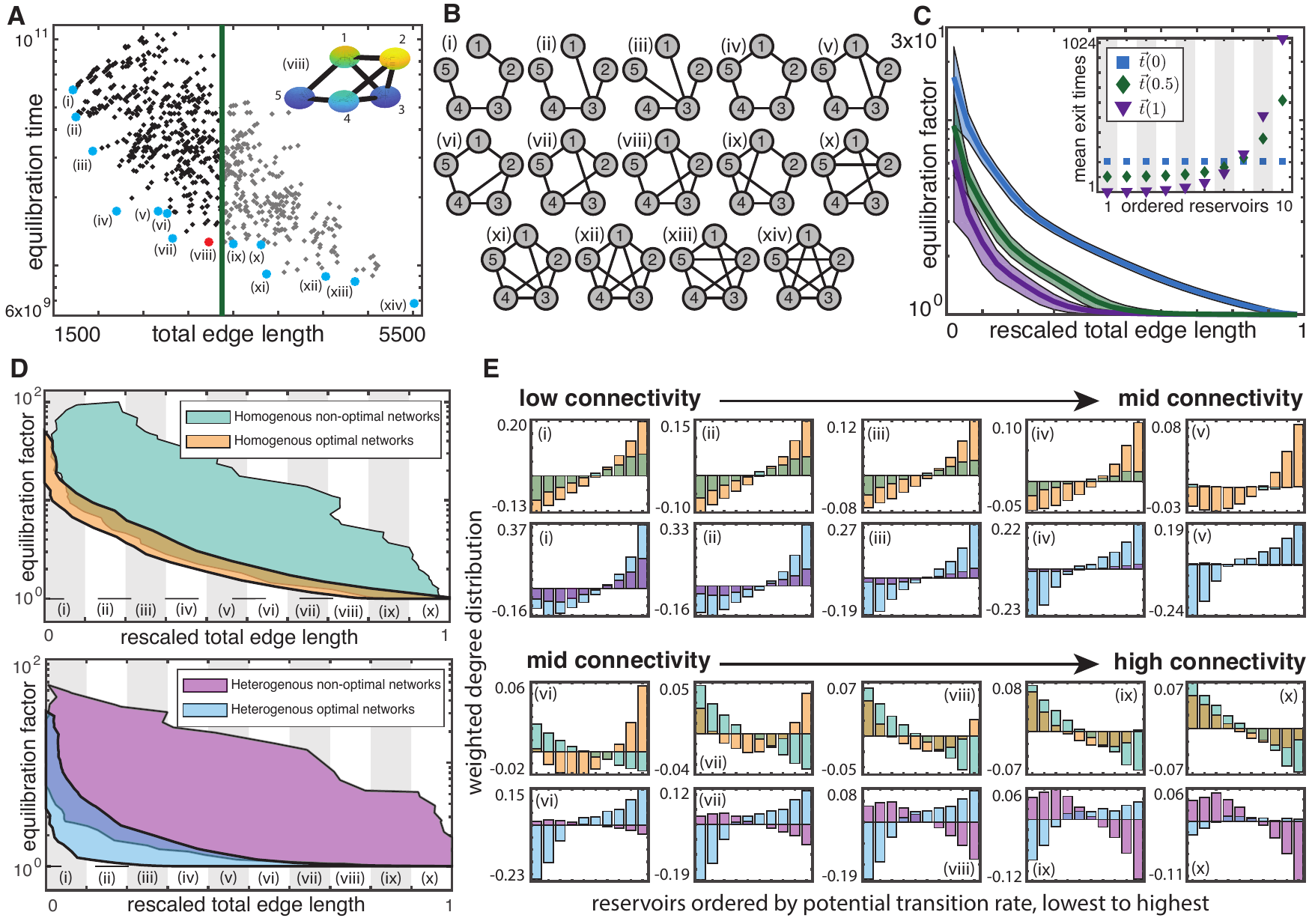}
\caption{Optimal topologies that minimise networked equilibration times. (A) Equilibration time and total edge length for all 728 connected topologies for a configuration of five reservoirs (inset). The optimal frontier is highlighted in asterisks and given labels (i)-(xiv). For a given restriction (vertical line) the optimal network is given by label (viii). (B) Topologies of the networks that lie on the optimal frontier in panel (A). (C) Averaged coordinates from numerically estimated optimal frontiers (SI Numerical Methods Section~$4.1$) with increasing levels of reservoir heterogeneity. The vectors of reservoir mean exit times are $\vec{t}(0)$, $\vec{t}(0.5)$ and $\vec{t}(1)$ (inset). The shaded regions represent the standard deviation either side of the mean. (D) The range of equilibration factors of both optimal and non-optimal networks for homogeneous, $\vec{t}(0)$, and heterogeneous, $\vec{t}(1)$, reservoir geometries. (E) The bar charts (i)-(x) show the weighted degree distribution for both optimal and non-optimal networks across the range of rescaled total edge lengths. The heterogeneous vector of reservoir mean exit times used in (C) and (E) is $\vec{\tau}= \left( 2^1, 2^2, \ldots , 2^{10}\right)$. All data presented in (C)--(E) uses the same 5000 configurations of 10 reservoirs with randomly generated ensembles of narrow channel lengths $\mathcal{K}$ (SI Numerical Methods Section~$4.3.1$).}
\label{Fig3}
\end{figure*}

\paragraph*{A global envelope of optimal networks}
To explore properties of the optimal frontier (Fig.~3A) we temporarily assume homogeneous reservoir exit times, an assumption that models environments with regular periodic structure such as synthetic porous nanomaterials \cite{Martin1994_Science}. Thus, networks that lie on the optimal frontier depend solely on the ensemble $\mathcal{K}$ of all possible narrow channel lengths which, in turn, depends upon the spatial configuration of the reservoirs (SI Numerical Methods Section~$4.3.1$). Comparing optimal frontiers between distinct ensembles $\mathcal{K}$ requires two rescalings (SI Optimal Networks Section~$2.1$). Firstly, the total edge length of a network must lie in an interval defined by the total edge lengths of the minimum spanning tree and the complete network. The rescaled total edge length linearly maps this interval to lie between zero and one and provides a dimensionless measure of connectedness. Such a rescaling ensures that the rescaled total edge length contains information about the distribution of all possible edge lengths not just the edge lengths present in the current network. Secondly, the equilibration times are normalised by the minimum equilibration time, which belongs to the complete network, and thus yields values greater than or equal to one which represent how many factors slower equilibration occurs for a given network compared to the complete network.

Numerical evidence strongly supports the hypothesis that the rescaled optimal frontier follows a global curve that is independent of $\mathcal{K}$ and hence is independent of the spatial configuration of the reservoirs (Fig.~S6). Over many distinct ensembles $\mathcal{K}$, the variation in the optimal frontier becomes vanishingly small as the number of reservoirs increases (Fig.~S6B--D). The global curve persists even for an ensemble of channel lengths that are intentionally sampled from an extremely heterogeneous distribution to ensure that the curve is not a feature of our sampling procedure (Fig.~S6E). The apparent persistence of the globally optimal frontier has significant implications for optimal network design; testing the optimality of a proposed network merely requires direct comparison of the rescaled total edge length and equilibration factor (two cheap-to-compute network statistics) to the global curve. Moreover, the global curve provides a benchmark to compare the efficacy of algorithms designed to efficiently construct optimal or close to optimal networks (SI Optimal Networks Section~$4.2$).

\paragraph*{Reservoir heterogeneity leads to minimised equilibration times in cases of restricted connectivity}

Complex geometries can exhibit a range of microscopic spatial structures (Fig.~1B), and this gives rise to  reservoir heterogeneity within the corresponding networks (Fig.~1C). The effects of such heterogeneities can be encapsulated by a vector of distinct reservoir exit times $\vec{\tau}$. To meaningfully compare the optimal frontiers that arise from two different vectors of reservoir exit times we require that both vectors have the same ensemble average $\langle \vec{\tau} \rangle$. Such a requirement guarantees that the narrow channel equilibrium occupancy is held constant (SI Models Section~$1.1.2$ Eq.~[13]), and thus changes in optimal equilibration times occur solely due to reservoir heterogeneity rather than a change in crowding effects\footnote[1]{The narrow channel equilibrium occupancy can be viewed as the probability that an attempted jump to a lattice site within a narrow channel is aborted, and thus quantifies the strength of the crowding effects.}. To systematically explore how heterogeneity impacts the optimal frontier we introduce a vector of reservoir exit times $\vec{t} \left( \phi \right)$ with $i$-th entry $t_i \left( \phi \right) = \left( 1 - \phi \right) \langle \vec{\tau} \rangle + \phi \tau_i$ for $\phi \in [0,1]$. \review{The new reservoir exit times $\vec{t} \left( \phi \right)$ depend upon a heterogeneous yet arbitrary vector of mean exit times $\vec{\tau}$, and} the parameter $\phi$ controls the extent of reservoir heterogeneity, where $\vec{t}(0) = \left( \langle \vec{\tau} \rangle , \ldots , \langle \vec{\tau} \rangle \right)$ represents homogeneous reservoirs,  ensuring $\langle \vec{t} \left( \phi \right) \rangle = \langle \vec{\tau} \rangle $ for all values of $\phi$. For increasing levels of reservoir heterogeneity, which is achieved by increasing $\phi$ (Fig.~3C, inset), the shape of the optimal frontier changes significantly, and optimal networks achieve globally minimised equilibration times (close to the equilibration time of the complete network) with significantly reduced connectivity (Fig.~3C). Thus, heterogeneity within the internal structure of complex environments has the potential to facilitate globally efficient equilibration in environments with spatially restricted connectivity.

\paragraph*{Optimal networks have distinct topological structure}

The topological structure of optimal networks can be characterised via a weighted degree distribution that arises from considering the diagonal entries of the weighted graph Laplacian, $\mathbf{F}^{\text{HD}}$. For the complete network the $i$-th diagonal entry of $\mathbf{F}^{\text{HD}}$ is proportional to $R_i = \sum_{j \neq i} \left( \tau_j K_{(i,j)} \right)^{-1}$ which represents the total transition rate out of the $i$-th reservoir when viewing the weighted graph Laplacian as a rate matrix (SI Optimal Networks Section~$2.4$). As every connection between reservoirs is present in the complete network we term $R_i$ the potential transition rate of reservoir $i$ and we order the reservoirs such that $R_1 < \cdots < R_{|\mathcal{V}|}$. The ratio of the $i$-th diagonal entry of $\mathbf{F}^{\text{HD}}$ for a network $\mathcal{G}$ with adjacency matrix $\mathcal{A}$ and the  $i$-th diagonal entry of $\mathbf{F}^{\text{HD}}$ for the complete network is $W_i \left( \mathcal{G};\mathcal{K} , \vec{\tau} \right) = \sum_{j \neq i} \mathcal{A}_{i,j} \left( \tau_j K_{(i,j)} \right)^{-1} / R_i$ and represents the fraction of the potential transition rate of reservoir $i$ present in the network $\mathcal{G}$. The weighted degree distribution is defined by $w_i \left( \mathcal{G};\mathcal{K} , \vec{\tau} \right)  = W_i \left( \mathcal{G};\mathcal{K} , \vec{\tau} \right) - \sum_{j=1}^{|\mathcal{V}|} W_j \left( \mathcal{G};\mathcal{K} , \vec{\tau} \right) / |\mathcal{V}|$ for $1 \leq i \leq |\mathcal{V}|$. The weights $w_i \left( \mathcal{G};\mathcal{K} ,\vec{\tau}\right)$ are translated such that if $w_i \left( \mathcal{G};\mathcal{K},\vec{\tau} \right) > 0$ then the $i$-th reservoir has a ratio $W_i \left( \mathcal{G};\mathcal{K},\vec{\tau} \right)$ greater than the network average, and is referred to as being over-represented in the network. Similarly, a reservoir with $w_i \left( \mathcal{G};\mathcal{K},\vec{\tau} \right) < 0$ is said to be under-represented. Naively, one might expect reservoirs with the highest potential transition rates to be over-represented in any network that lies on an optimal frontier, as connections between reservoirs with high transition rates should encourage faster equilibration. However, we discover that the structure of optimal networks varies greatly depending on the restriction on the rescaled total edge length.
 
The structure of networks that lie on the optimal frontier compared with non-optimal networks (randomly sampled connected topologies) is distinct across all rescaled total edge lengths (Fig.~3D), with the greatest contrast seen for networks with mid-range connectivity (Fig.~3D,E (iii)-(viii)). Moreover, the weighted degree distribution highlights how reservoirs can be connected to achieve optimality. Optimal networks with low connectivity and reservoir homogeneity prefer connections between reservoirs with high potential transition rates (Fig.~3E(i) orange/green), on the other hand, for highly connected optimal networks the relative importance of the reservoirs is reversed (Fig.~3E(x) orange/green). Interestingly, as connectivity varies between the two extremes, optimal networks involve over-representation of reservoirs with both high and low potential transition rates (Fig.~3E,(vi)—(viii) orange/green). This transitional behaviour vanishes if reservoir heterogeneity is sufficiently high, where reservoirs with high potential transition rates are over-represented for all levels of connectivity (Fig.~3E blue/purple). Even for highly connected networks, the absence of a single important connection between reservoirs can significantly increase equilibration times by over a factor of four in heterogeneous environments (SI Section~$2.5$, Fig.~S8). Collectively our results demonstrate the ability of the weighted degree distribution, as well as the graph Laplacian $\mathbf{F}^{\text{HD}}$, to reveal connections between geometric structure and optimal transport that could not have been identified with traditional modelling approaches.

\paragraph*{The dynamics of tagged individuals are highly-sensitive to narrow channel length}

The detailed dynamics of a tagged individual within a population is of interest across a broad range of disciplines \cite{WeiScience2000,Jaqaman2008_NatMeth}. For example, the differentiated fate of a stem cell can hinge upon the spatial and temporal dynamics of a single protein within the crowded intracellular environment \cite{Androutsellis-Theotokis2006_Nature}. A significant benefit of our framework is that it readily extends to provide information at the level of a single tagged individual. The microscopic dynamics of a tagged individual are identical to the dynamics of any individual within the FMM. Therefore, the transition of a tagged individual between adjacent reservoirs $i$ and $j$ via a connecting narrow channel of length $K_{(i,j)} $ occurs as follows. First a tagged individual in reservoir $i$ will jump into the first lattice site adjacent to reservoir $i$ on the narrow channel connecting the $i$-th and $j$-th reservoirs. In the high-density regime, the tagged particle will jump back into reservoir $i$ at rate $\alpha$ or jump into the adjacent (second) lattice site when a background individual is exchanged from the $i$-th to the $j$-th reservoirs. The latter jump occurs at a significantly lower rate $k_{i,j}^{\text{HD}} \left( \vec{n} \right)$ (SI Eq.~[16]) and almost always the tagged individual returns to the $i$-th reservoir.  For this reason we consider a tagged individual to have `properly' entered the narrow channel only when it reaches the second lattice site (Fig.~4A(i),(ii)). Then, as background individuals continue to be exchanged between reservoirs $i$ and $j$, the tagged individual undergoes a random walk along the sites of the narrow channel until either being absorbed back into the $i$-th reservoir (Fig.~4A(i)), or absorbed into the $j$-th reservoir (Fig.~4A(iii)). The latter occurs with probability $p_{\text{TI}} \left( x_i, x_j; K_{(i,j)} \right)$, where the fractions of the population that occupy the $i$-th and $j$-th reservoirs at the moment when the tagged individual first enters the narrow channel (Fig.~4A(ii)) are denoted $x_i$ and $x_j$, respectively. The probability $p_{\text{TI}}\left( x_i, x_j; K_{(i,j)} \right)$ is given by 
\begin{equation}\label{eq:HP}
p_{\text{TI}} \left( x_i,x_j;K_{(i,j)}\right) \sim f_0 \left( x_i, x_j \right) \left[ \sum_{k=0}^{K_{(i,j)}-2}  f_k \left( x_i, x_j \right)\right]^{-1},
\end{equation}
where the functions $ f_k \left( x_i, x_j \right)$ are given by
\begin{equation}\label{eq:f}
 f_k \left( x_i, x_j \right) = \left( \dfrac{\tau_i}{\tau_j} \right)^k \left( \dfrac{x_i}{x_i+x_j} + \dfrac{2-k}{N_{\text{HD}}\left( x_i + x_j \right)} \right)^{N_{\text{HD}}x_i + 3/2 - k} \left( \dfrac{x_j}{x_i+x_j} + \dfrac{1+k}{N_{\text{HD}}\left( x_i + x_j \right)}\right)^{N_{\text{HD}}x_j + 1/2 + k}.
\end{equation}
The probability $p_{\text{TI}}$ is referred to as the \textit{tagged individual crossing probability} (see SI Models Section~$1.4.1$, Eqs.~[S33-S40] for a derivation).

\begin{figure}[tb]
\centering
\includegraphics[width=12.085cm]{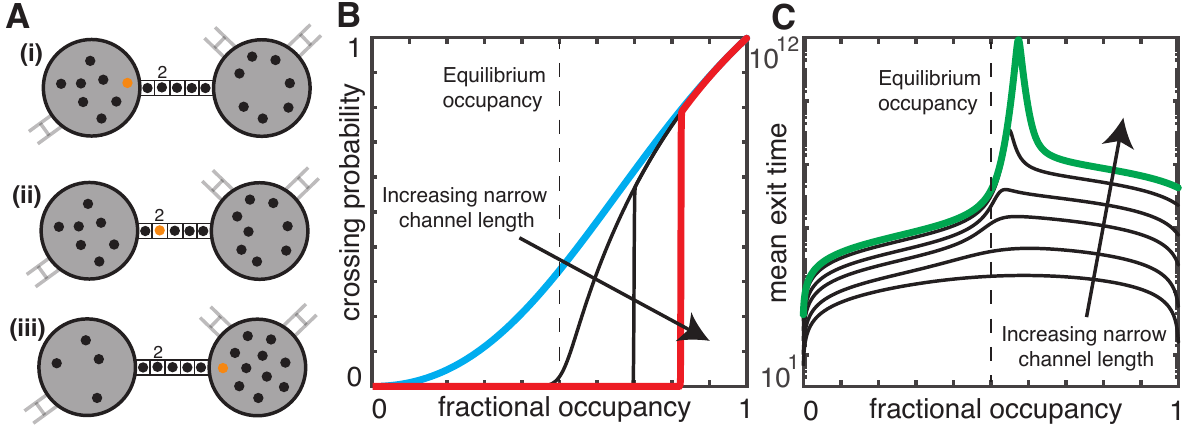}
\caption{Narrow channel lengths affect tagged individual dynamics. (A) Diagrammatic representation of how a tagged individual (orange) transitions from one reservoir to another, and the net exchange of background individuals (black) required to do so. (B) The tagged individual crossing probability $p_{\text{TI}}$ (SI Models Section~$1.4.1$, Eq.~[S39]) for increasing narrow channel lengths $K_{(i,j)} \in \left\lbrace 4,25,400,650 \right\rbrace$. The blue and red lines correspond to channel lengths of four and $650$, respectively. (C) The tagged individual mean exit time $m_{\text{TI}}$ (SI Models Section~$1.4.2$ Eq.~[S42]) for increasing narrow channel lengths $K_{(i,j)} \in \left\lbrace 5,10,25,50,100,150\right\rbrace$ where the curve corresponding to $K_{(i,j)} =150$ is highlighted in green. The parameters used in (B) and (C) are $\tau_i=\tau_j=0.1$ and $N=10^3$.}
\label{Fig4}
\end{figure}

The successful crossing of a tagged individual that has just entered the narrow channel (Fig.~4A(ii)) requires a net exchange of several background individuals from the $i$-th to the $j$-th reservoir (SI Models Section~$1.4.1$). The probability, $p_{\text{TI}}$, that this net exchange occurs depends on $x_i/ \left( x_i + x_j \right)$, the fractional occupancy of the $i$-th reservoir relative to the $j$-th (Fig.~4B). For fractional occupancies greater than the equilibrium fractional occupancy, which is $x_i^*/ \left( x_i^* + x_j^* \right)$ where $x_i^* = \tau_i / \sum_{j=1}^{|\mathcal{V}|} \tau_j $ (Fig.~4B, vertical line), as the population equilibrates there is a bias favouring exchange of background individuals from the $i$-th to the $j$-th reservoir which subsequently increases the tagged individual crossing probability. However, the probability $p_{\text{TI}}$ becomes incredibly sensitive to the fractional occupancy as the length of the narrow channel increases (Fig.~4B, arrow). Indeed, a successful crossing of a tagged individual can require that the fractional occupancy of the $i$-th reservoir is significantly above the equilibrium occupancy (Fig.~4B, red curve). Thus, in an equilibrated population, the probability that a tagged individual traverses between two adjacent reservoirs is effectively zero if the narrow channel is too long\footnote[3]{ For homogeneous reservoirs ($\tau_i = \tau$) we find that the tagged individual crossing probability is effectively zero when $K_{(i,j)} \geq \sqrt{N/|\mathcal{V}|}$ (SI Models Section~$1.4.4$, Eq.~[S49] and Fig.~S5).}.

The temporal dynamics of tagged individuals are drastically affected by narrow channel length. For an individual that has just entered the second lattice site along the narrow channel (Fig.~4A(ii)), the tagged individual mean exit time, $m_{\text{TI}} \left( x_i, x_j; K_{(i,j)} \right)$, denotes the mean time taken for the tagged individual to exit the channel at either end, and is given by
\begin{equation}\label{eq:MFPT}
m_{\text{TI}} \left( x_i , x_j ; K_{(i,j)} \right) \sim p_{\text{TI}} \left( x_i , x_j ; K_{(i,j)} \right) \left[ h_1 \left( x_j \right) + \sum_{k=2}^{K_{(i,j)}-2} \sum_{\ell=1}^k \dfrac{h_{\ell}(x_j) f_k \left( x_i, x_j \right)}{f_{\ell} \left( x_i, x_j \right)}\right],
\end{equation}
where $h_{\ell}(x_j) = \left( K_{(i,j)}-1\right) \left( N_{\text{HD}} x_j + \ell \right) / \left( \tau_j \alpha^2 \right)$, and the $ f_k \left( x_i, x_j \right)$ are as in Eq.~\eqref{eq:f} (see SI Models Section~$1.4.2$ Eqs.~[S41,S42] for a derivation). The effect of increasing the length of the narrow channel $K_{(i,j)}$ on $m_{\text{TI}} $ is two-fold. Firstly, the tagged individual mean exit time increases drastically (Fig.~4C, arrow). Secondly, as for $p_{\text{TI}} $, $m_{\text{TI}} $ becomes incredibly sensitive to the fractional occupancy of the two reservoirs. Fractional occupancies of the $i$-th reservoir slightly above equilibrium introduce a temporary bias that encourages the tagged individual to move further along the channel. Once the background individuals have equilibrated the bias is removed whilst the tagged individual remains within the internal lattice sites of a long narrow channel. Relying solely on the unbiased stochastic fluctuations of background individuals increases the time taken for the tagged individual to exit the channel by many orders of magnitude (Fig.~4C, peak of green curve), in particular this time can significantly exceed the equilibration time of the entire population (Fig.~S3B).

\paragraph*{Crowding alters the paths taken by tagged individuals}

Exploration of the motion of tagged individuals throughout a complex and crowded environment requires highly-scalable networked transport models capable of extracting information at the level of the individual. We extend our framework, using the concepts of the tagged individual crossing probability, $p_{\text{TI}}$, and mean exit time, $m_{\text{TI}}$, to provide such models of individual dynamics within both fixed and stochastically fluctuating background populations (SI Models Section~$1.4.3$ and $1.4.4$). For a fixed equilibrated population, the spatial dynamics of a tagged individual can be investigated via a discrete networked random walk model. The dynamics of this random walk model are revealed via analysis of the transition matrix, rather than via stochastic simulation (SI Numerical Methods Section~$4.5$). As such, many useful statistics describing the dynamics of tagged individuals within complex and crowded environments are immediately available. To highlight the utility of the discrete random walk model we consider a first-passage process \cite{Redner2001_Book}, reminiscent of the Notch signalling pathway where an intracellular protein traverses between the cellular and nuclear membranes \cite{Androutsellis-Theotokis2006_Nature}. We adopt a caricature representation of the intracellular geometry in the form of a random geometric network (Fig.5A,B and SI Numerical Methods Section~$4.3.2$), and consider the first-passage properties of a tagged individual initially within a reservoir adjacent to the cellular membrane (Fig.~5A,B, outer circle) whose position evolves according to the networked discrete random walk (SI Numerical Methods Section~$4.5$) before terminating at a reservoir adjacent to the nucleus (Fig.~5A,B, inner circle). 

By comparing our discrete random walk model to an almost identical model that does not consider crowding effects (SI Models Section~$1.4.4$, Eqs.~[S44,S45]), we discover that crowding effects within the narrow channels drastically alters the paths taken by tagged individuals (Fig.~5). The expected number of times an individual traverses a narrow channel becomes highly sensitive to local network topology when crowding effects are incorporated (Fig.~5A, B). The negligible chance that a crowded tagged individual traverses a long narrow channel (Fig.~4B, Fig.~S6) significantly widens the distribution of the expected number of crossings (Fig.~5C and A,B highlighted regions), and tagged individuals follow paths that visit reservoirs connected via short narrow channels significantly more often than longer channels (Fig.~S4). Favouring shorter narrow channels subsequently favours indirect paths (Fig.~S4A) resulting in an increase in the total path length of an individual that moves from the cellular to the nuclear membrane (Fig.~5D). The alterations of the dynamics of tagged individuals due to crowding, as detailed above, has important implications for the efficiency of signalling pathways and subsequent downstream processes \cite{Mukhopadhyay2013_PLoSCB}.

\begin{figure}[tb]
\centering
\includegraphics[width=\columnwidth]{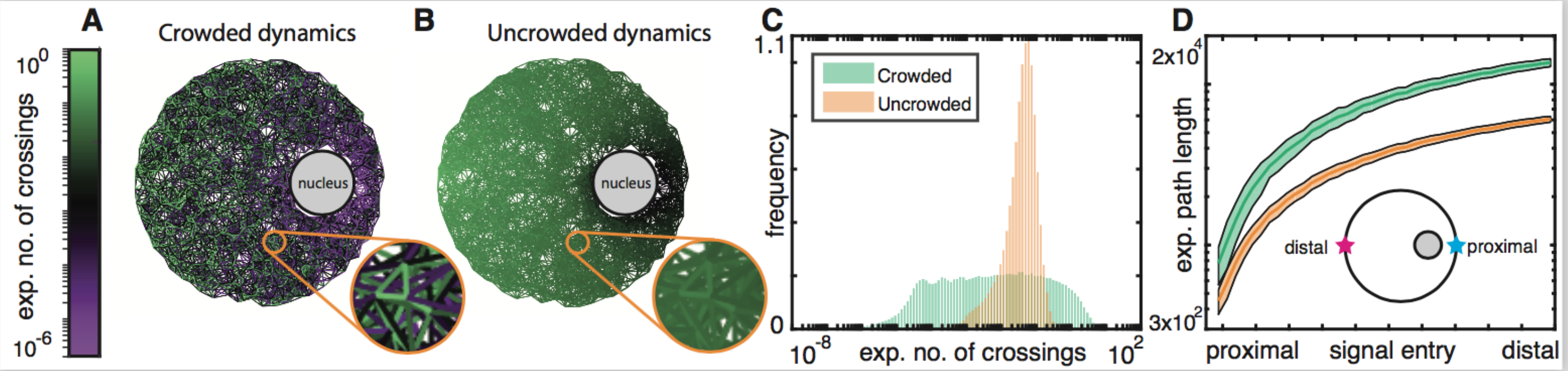}
\caption{The paths of tagged individuals between the cellular and nuclear membranes are heavily altered due to crowding. (A-B) Expected number of crossings of each narrow channel with and without crowding in a realisation of a cellular signalling network with 1000 reservoirs (SI Numerical Methods Section~$4.5$). (C) The normalised frequency of the number of channels for a given expected number of crossings. (D)  Expected path length of a tagged individual as a function of the position of the initial entry at the cellular membrane. All data presented in (C-D) is averaged over 100 realisations of a random geometric network (SI Numerical Methods Section~$4.3.2$). Parameters for the discrete random walks (SI Models Section~$1.4.4$) used in (A-D) are $N=3 \times 10^4$, $\alpha=1$, and $\tau_i=0.1$ for every reservoir.}
\label{Fig5}
\end{figure}

\section*{Discussion}


Our results highlight how networks provide an effective framework through which to reveal and quantify the combined influences of geometry and crowding on the transport properties of individuals confined within finite complex environments. Moreover, detailed network analysis uncovered global relationships between population-level transport behaviour and optimal networked topologies, as well as the salient network features responsible for optimisation. Our framework provides new insights into the interplay between geometry, crowding and transport, with such insights being of relevance for a range of geometrically-regulated transport processes.


The work presented in this paper focusses on modelling the temporal fluctuations of particles within the reservoirs of a network. This was achieved through averaging the effects of crowding within narrow channels by assuming that the narrow channel occupancy is held constant. However, previous work has shown that crowding effects can give rise to strongly fluctuating densities within narrow channels \cite{Derrida2004}. As such, it would be interesting to extend our framework to allow for both the occupancy of reservoirs and narrow channels to fluctuate and explore whether our conclusions about optimal networked topologies change. This would offer a means to test whether the global envelope of optimal networks persists, providing further evidence of a possible universal relationship between rescaled total edge length and optimal equilibration.

Beyond identifying global connections between crowding, geometry and transport, our framework provides a highly efficient computational tool to perform more focussed investigations. For example, cardiomyocyte cells have an intracellular environment consisting of mitochondrial and myofibril filaments. Patients suffering from diabetic cardiomyopathy have been shown to have highly clustered and disordered mitochondrial distributions \cite{Jarosz2017_AJPCP,Ghosh2018_PLoSCB}. It has been hypothesised that these alterations to the intracellular geometry help regulate the transport of essential metabolites and increase the overall energy supply to the cell in an effort to maintain regular heart function \cite{Jarosz2017_AJPCP}. A networked modelling approach would offer a sophisticated investigation into the functional role of the observed intracellular restructuring. Analysing the structure of networks extracted from light-sheet microscopy images (SI Fig.~S$10$) would quantify the structural differences between healthy and diabetic cardiomyocytes and studying the transport of metabolites on these networks would provide a means to test whether these changes in intracellular geometry can improve cellular bionergetics. Moreover, the computational efficiency seen within our framework significantly increases the number of experimental images that can be studied compared with traditional approaches that are constrained by using high-resolution meshes. Studying a large number of distinct cellular geometries will be critical to developing a deeper understanding of how intracellular restructuring may regulate cellular bioenergetics.



Our framework offers numerous opportunities for generalisation. The RMM is formulated as a chemical reaction network \cite{Warne2019_JRSI} and thus can immediately support reactions between individuals, where subsequent coarse-graining will result in a Fokker-Planck equation capable of investigating geometry-controlled kinetics \cite{Benichou2010_NatChem}. Included within the SI is a generalisation of our framework that supports active transport (SI Extensions Section~$3.2$). We study a Partially Asymmetric Simple Exclusion Process (PASEP) that allows individuals to undergo bi-directional motion along narrow channels but with a bias in one direction. Extending to active transport opens the door to investigating the geometric influences on highly crowded transport phenomena across a wide array of spatial scales, from mRNA translocation along intracellular microtubule networks \cite{Hirokawa2009_NatRev_MCB}, to molecular trafficking between cells connected via cytonemes \cite{Teimouri2016_JPCL} or plasmodesmata \cite{Ueki2005_PNAS}, to the transport of sediment subjected to flows within porous media \cite{McDowell-Boyer1986_WWR}. 

Through reconceptualising how we model crowded, geometrically-constrained transport, we stand to gain significant new insights within fields such as optimal synthetic design \cite{Lebiedz2012_MMB} and molecular cell biology \cite{Baltimore1986_MCB}, to name a few. Furthermore, this work presents a versatile framework that paves the way for furthering our understanding of the fundamental connections between crowding, networked geometry and transport, beyond what has already been achieved under more traditional modelling paradigms.


\bibliography{scibib}

\section*{Acknowledgements}

The authors would like to thank Dr. Rajagopal for enlightening discussions and providing the imaging data used in Fig.~1A. The authors would like to additionally thank the anonymous reviewers for their comprehensive and constructive feedback. This work was supported by the EPSRC Systems Biology DTC Grant No. EP/G03706X/1 (D.B.W.), a Royal Society Wolfson Research Merit Award (R.E.B.), a Leverhulme Research Fellowship (R.E.B.), the BBSRC UK Multi-Scale Biology Network Grant No. BB/M025888/1 (R.E.B. and F.G.W.), and the Royal Society International Exchanges Scheme (F.G.W. and M.J.S.).

\section*{Competing Interests}
The authors declare no competing interests.

\section*{Author Contributions}
All authors contributed at all stages of this work.

\end{document}